\documentclass{aa}
\usepackage{txfonts}
\usepackage{epsfig}
\usepackage{subfigure}

\usepackage{times}
\usepackage{natbib}
\bibpunct{(}{)}{;}{a}{}{,}

\begin{document}

\title{The recent X--ray history of NGC~5506}
\author{S. Bianchi\inst{1}, I. Balestra\inst{1}, G. Matt\inst{1}, M.
Guainazzi\inst{2}, G.C. Perola\inst{1}}

\offprints{Stefano Bianchi\\ \email{bianchi@fis.uniroma3.it}}

\institute{Dipartimento di Fisica, Universit\`a degli Studi Roma Tre, Italy
\and XMM-Newton Science Operation Center/RSSD-ESA, Villafranca del Castillo,
Spain }

\date{Received / Accepted}

\authorrunning{S. Bianchi et al.}

\abstract{We present a detailed discussion of the spectral and spatial components
of NGC~5506,  based on XMM--$Newton$, $Chandra$ and BeppoSAX observations.
The overall picture consists of a nucleus absorbed by cold gas with column
density of $\approx10^{22}$ cm$^{-2}$ and surrounded by a Compton-thick torus,
whose existence is inferred by a cold reflection component and an iron K$\alpha$
line. On a much larger scale, a photoionized gas extended on $\approx350$ pc
reprocesses the nuclear radiation, producing a soft excess and ionized iron
lines. Noteworthy, we find no evidence for the presence of the accretion disc.


}

\maketitle

\section{Introduction}

NGC~5506 hosts a nearby (z=0.006) AGN, rather bright in hard X-rays. It has
been generally classified as a NELG. Recently \citet{nag02} observed the source in
the near--IR and discovered an heavily reddened (A$_V\sim$5) Narrow Line
Seyfert 1 nucleus.

The source, being very bright, was observed by all X--ray satellites, starting
with $Uhuru$. In recent times, it was observed by $Einstein$ \citep{mac82},
$EXOSAT$ \citep{pounds89} and $Ginga$ \citep{bmy93}; the latter fitted the
spectrum with a power law absorbed by neutral matter with column density of
$3-4\times10^{22}$ cm$^{-2}$, plus a soft excess. Furthermore, the data showed a
reflection component and an iron K$\alpha$ line \citep{bmy93}. Later on, $ROSAT$
HRI images suggested that the soft emission is extended and coincident with the
radio emission \citep{colbert98}. Results from $ASCA$ \citep{wang99}, BeppoSAX
\citep{per02} and an $RXTE$ variability analysis \citep{lamer00} pointed out the
complexity of the iron line profile, which was finally resolved by XMM-$Newton$
in at least two components, the narrower at 6.4 keV being likely associated with
the reflection component, arising from a neutral Compton-thick material
\citep{Matt01}. The origin of the bluer and broader component is less clear, but
may be associated to the soft excess (see Sect. \ref{softpar}). Finally, the
source is variable on short time scales, but no long term trend has been found
yet \citep[e.g.][]{papa02}.

In this paper we present results from two XMM--$Newton$ observations (the
first one simultaneous with a BeppoSAX observation, see \citealt{Matt01}) and a
$Chandra$/HETG observation. We have also reanalysed past BeppoSAX and ASCA
observations (see Table~\ref{log}). All these observations will allow us to
check and refine the interpretation proposed by \citet{Matt01} and \citet{bm02}, in
which the nucleus is surrounded by at least two reflecting regions, one
Compton--thick and neutral and the other one Compton--thin and highly ionized,
and obscured by  a Compton--thin cold absorber.

We will assume H$_0=75$ km/s/Mpc throughout the paper.

\section{\label{data}Observations and data reduction}

\begin{table}[t]

\caption{\label{log}The log of all analysed observations, and 
exposure times}

\begin{center}

\begin{tabular}{cccc}
\hline \textbf{Date} & \textbf{Mission} & \textbf{Instr.} & \textbf{T$_{exp}$
(ks)} \\
\hline 01/30/1997 & $ASCA$   & \textsc{sis0-1} & 39\\
\hline
            &          & \textsc{lecs} & 16\\
 01/30/1997 & BeppoSAX & \textsc{mecs} & 39\\
            &          & \textsc{pds} & 17\\
\hline
            &          & \textsc{lecs} & 9\\
 01/14/1998 & BeppoSAX & \textsc{mecs} & 39\\
            &          & \textsc{pds} & 17\\
\hline
 12/31/2000 & $Chandra$ & \textsc{acis-s hetg} & 90\\
\hline
            &          & \textsc{lecs} & 29\\
 02/01/2001 & BeppoSAX & \textsc{mecs} & 78\\
            &          & \textsc{pds} & 38\\
\hline
 02/02/2001 & XMM-$Newton$ & \textsc{epic pn}  & 14 \\
\hline
 01/09/2002 & XMM-$Newton$ & \textsc{epic pn}  & 10 \\
\hline
\end{tabular}

\end{center}

\end{table}

\subsection{XMM-Newton}

NGC~5506 was observed twice by XMM-$Newton$, on February 2001 and on January
2002 (Table \ref{log}). We defer the reader to \citet{Matt01} for details on the
first observation and the related data reduction. Both observations were
performed with the imaging CCD cameras, the EPIC-MOS \citep{turner01} and the
EPIC-pn \citep{struder01} operating in Large Window mode and the Medium filter.
X-ray events corresponding to pattern 0 were used for the pn and 0-12 for the
MOS. We will not deal with the RGS spectra in this paper, because of the too
poor statistics. Data were reduced with SAS 5.3.0, including the first
observation which was reprocessed. As an effect of the new adopted
response matrix, we found a slightly different normalization factor
between PDS and pn, 1.15, instead of 1.215 used by \citet{Matt01}. EPIC-pn
spectra (0.5-10 keV) and lightcurves were extracted from a radius of
$40\arcsec$. An extraction region of $45\arcsec$ radii was instead adopted for
the EPIC MOS1 and MOS2 spectra (0.3-10 keV). Both observations are affected by
negligible pileup (less than 1\% in the pn). Spectra were analysed with
\textsc{Xspec} 11.1.0.

\subsection{\label{chandra}Chandra}

The $Chandra$ observation was performed on New Year's Eve 2000 (Table \ref{log})
with the Advanced CCD Imaging Spectrometer (ACIS-S: Garmire et al., in
preparation) and the High-Energy Transmission Grating Spectrometer (HETGS:
Canizares et al., in preparation) in place. The high flux of the source
together with the default frame time of 3.2 s resulted in a 0th order spectrum
which is strongly affected by pileup (70\% according to
WebPIMMS\footnote{http://heasarc.gsfc.nasa.gov/Tools/w3pimms.html}). Therefore,
we will only use the 1st order co-added MEG and HEG spectra in this paper. Data
were reduced with the Chandra Interactive Analysis of Observations software
(CIAO 2.2.1), using the Chandra Calibration Database (CALDB 2.10). Grating
spectra were analysed with $Sherpa$ 2.2.1.

\subsection{BeppoSAX}

BeppoSAX observed the source three times, on January 1997, January 1998 and
February 2001 (Table \ref{log}). The first two observations were published by \citet{risa02} and
\citet{per02}, while the third one by \citet{Matt01}, together with the
simultaneous XMM-Newton observation. Data reduction followed the standard
procedure presented by \citet{guainazzisax99}, using HEAsoft 5.1. Spectra were
extracted from regions of radius $8\arcmin$ for the LECS and $4\arcmin$ for the
MECS, and analysed with \textsc{Xspec} 11.1.0. A normalization factor of 0.86
was adopted between the PDS and the MECS \citep{fiore99}, appropriate for PDS
spectra extracted with fixed rise time threshold.

\subsection{ASCA}

The source was observed by $ASCA$ on January 1997, simultaneously with the
first BeppoSAX observation. Data were published by \citet{wang99}. SIS spectra
(0.5-10 keV) were downloaded from the ASCA Archive and background was extracted
from blank field observations of SIS0 e SIS1. Data were analysed with HEAsoft
5.1. and \textsc{Xspec} 11.1.0. Following the procedure adopted for the
simultaneous XMM-BeppoSAX observation, we adopted a normalization factor of
0.898 between the PDS and the SIS0 for the simultaneous $ASCA$-BeppoSAX
observation.\\

In the following, errors are at the 90\% confidence level for one interesting
parameter ($\Delta \chi^2 =2.71$).

\section{Data analysis}

\subsection{Flux and spectral variability}

\begin{figure}

\begin{center}

\subfigure[\label{hardratio}]
{\epsfig{figure=hardratio.ps,width=5.5cm,angle=-90}}
\subfigure[\label{flux}]
{\epsfig{figure=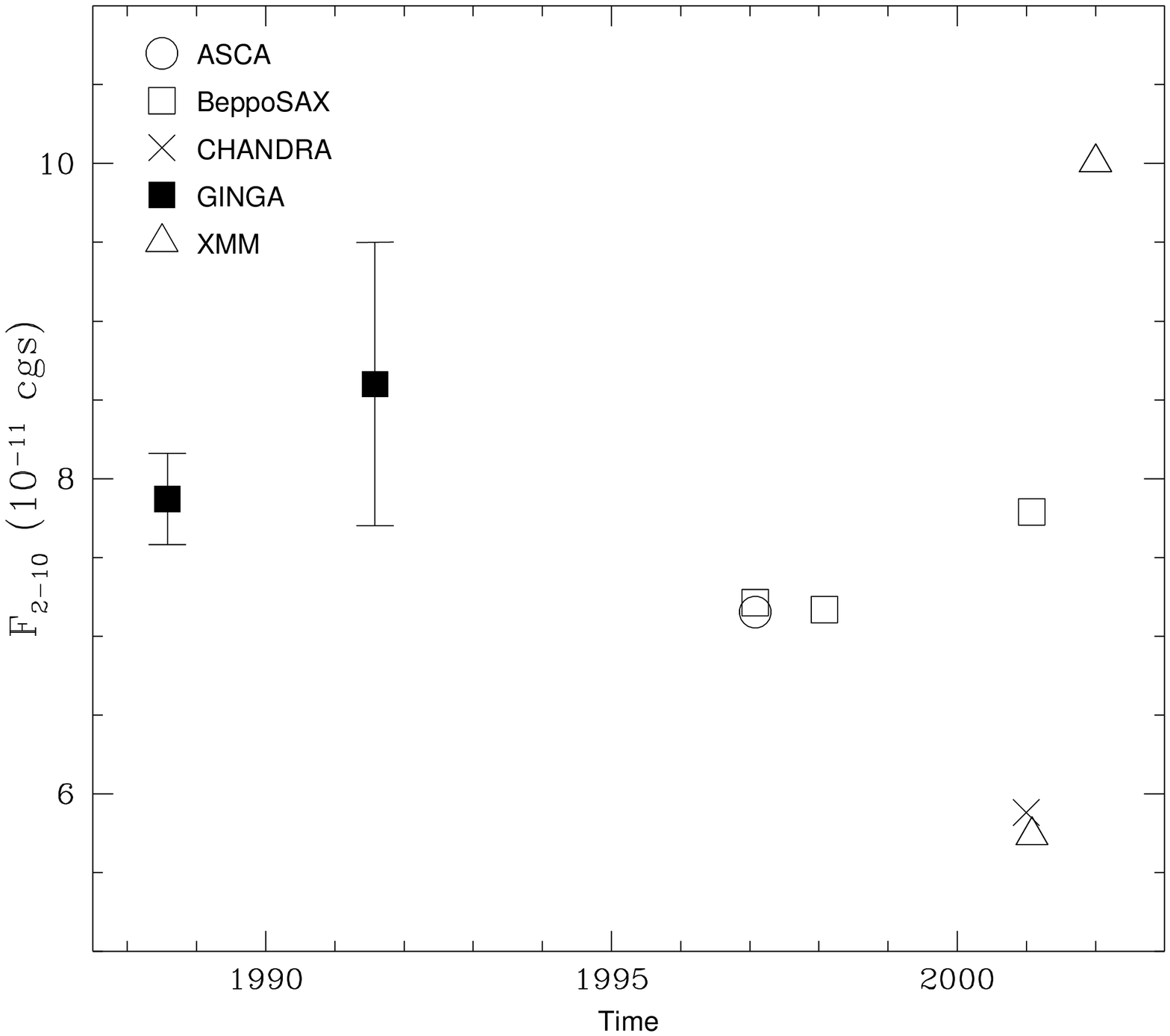,width=5.5cm}}
\end{center}

\caption{(a) Lightcurve and hardness ratio for the second
XMM-$Newton$ observation. (b) Flux (2-10 keV) of NGC~5506 for all the
observations analysed in this paper (see Table \ref{log} for details). The two
GINGA points were taken from \citet{bmy93} and \citet{np94}}

\end{figure}

We analysed a total of seven X-ray observations of NGC~5506, covering a history
of five years (see Table \ref{log}). The source shows a remarkable variability both
during each observation (see Fig. \ref{hardratio} for an example), and between
different observations (Fig. \ref{flux}). However, the large variation between
the two XMM observations may not be related to a real long-term variability,
because these observations are rather short. Therefore, the measured fluxes
could be the effect of a large amplitude short-term variability, with the
observations sampling different phases of it. This is also supported by the
third long BeppoSAX observation, which has a average flux much higher than the
first XMM-$Newton$ one, which covers only a small fraction of it. Moreover, the
three BeppoSAX observations have a remarkably similar average flux over a period
of time of 4 years, suggesting that any variability time scale is smaller than
about 1-2 days, i.e. the typical elapsed time of the BeppoSAX pointings.

\begin{figure}

\begin{center}

\subfigure[\label{nH}]
{\epsfig{figure=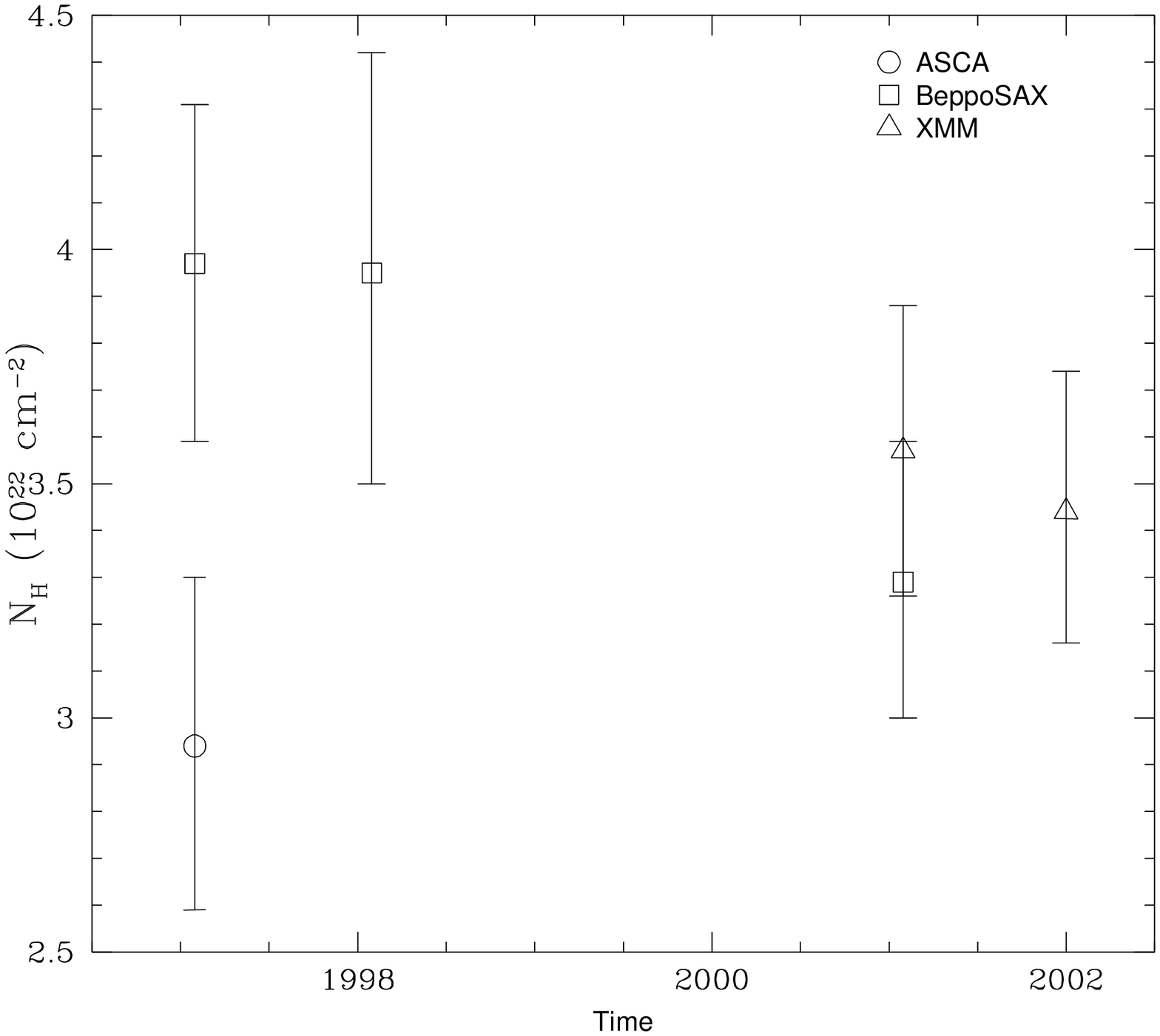,width=5.5cm}}
\subfigure[\label{gamma}]
{\epsfig{figure=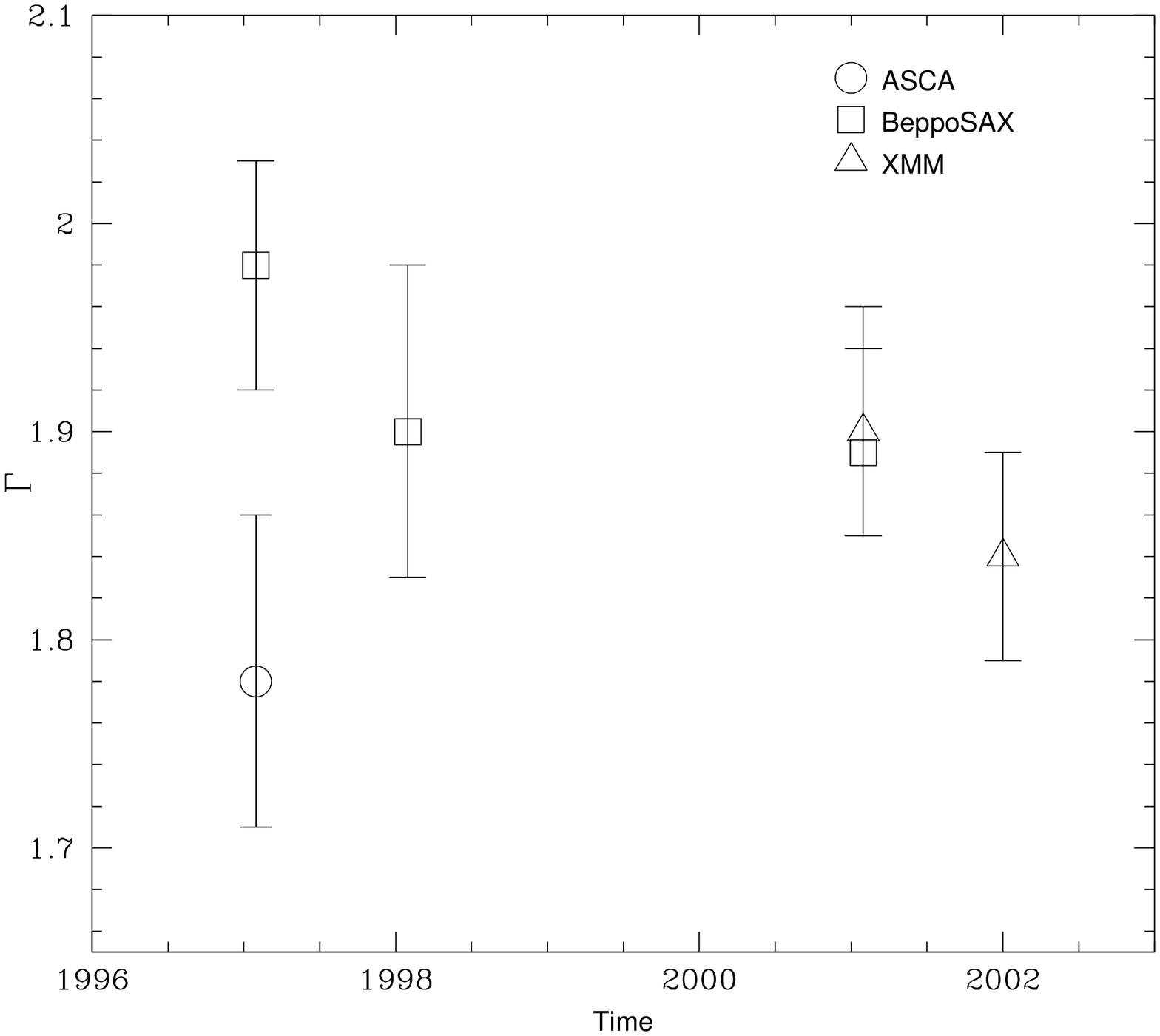,width=5.5cm}}

\end{center}

\caption{Column density (a) and photon index (b) of NGC~5506 for the selected
observations, adopting a simple absorbed power law model plus iron line in the
2.5-10 keV band (see text for details).}

\end{figure}

The hardness ratio in each observation seems to indicate that there is no
significant short-term spectral variation (Fig. \ref{hardratio}). On the other
hand, in order to understand if the long-term flux variability is associated to
spectral variations of the nuclear continuum, we performed a fit with a simple
absorbed power law to all the analysed observations, with the only exception of
$Chandra$, because the 0th order spectrum is largely affected by pileup (see
Sect. \ref{chandra}), while in the 1st order spectra there are not enough
continuum photons.  We restricted our fits to the band 2.5-10 keV, in order to
avoid the soft excess and the reflection component (we will deal with their
temporal behaviour later). The results are shown in Fig. \ref{nH} and
\ref{gamma}: despite the large flux variability, no significant variations are
present in the spectral shape of NGC~5506. Only the $ASCA$ parameters (fully
compatible with those derived by \citealt{wang99}) are systematically lower than
the other ones but this is probably due to calibration problems, being
inconsistent with the simultaneous BeppoSAX observation.

\subsection{Spectral analysis}

\begin{figure}
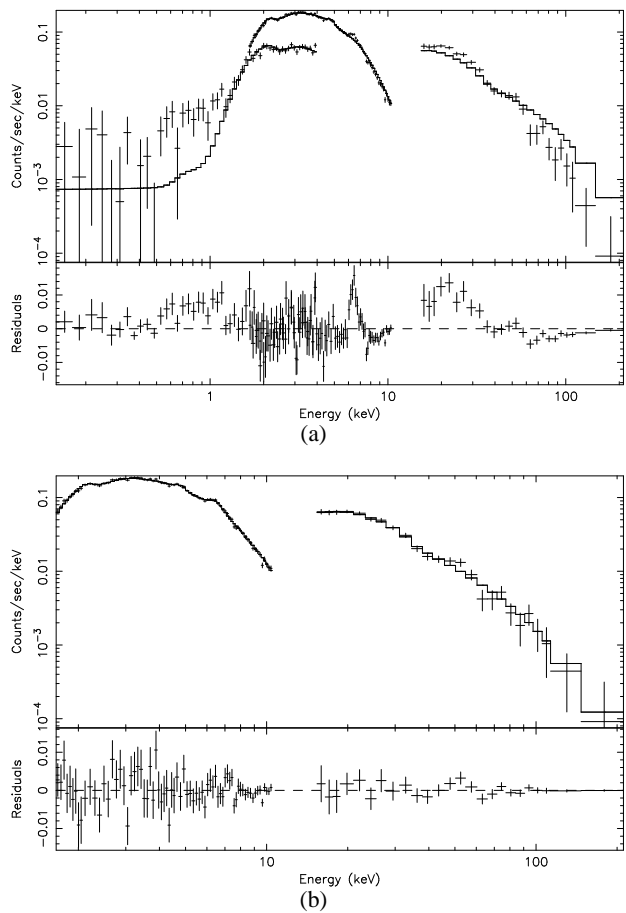


\begin{center}

\subfigure[\label{saxpo}]
{\epsfig{figure=saxpo.ps,width=5.5cm,angle=-90}}
\subfigure[\label{saxpexrav}]
{\epsfig{figure=saxpexrav.ps,width=5.5cm,angle=-90}}

\end{center}

\caption{(a) Spectra and residuals from LECS, MECS and PDS from the last
BeppoSAX observation when fitted with a simple absorbed power law. (b) Spectra
and residuals from MECS and PDS from the last BeppoSAX observation when a
reflection component model plus iron line model is adopted (see text for
details).}

\end{figure}

We start our spectral analysis from the three BeppoSAX observations, which give
valuable informations on the broadband spectrum of the source. If we adopt a
simple model with a power law absorbed by a column density of neutral gas, we
get unacceptable fits (reduced $\chi^2$ higher than 2) for all the three
datasets. Three features clearly arise from the residuals: a soft excess, an
iron line and an excess at high energy, likely a signature of Compton reflection
(see Fig. \ref{saxpo} for the third observation).

We note that the soft excess is an important component only in the LECS
band and clearly does not affect the spectrum at higher energies. Therefore, we
will exclude from our first analysis the LECS data and focus our attention in
the next section on the reflection component and the iron line. The same
reason led us to ignore all data below 2.5 keV in the pn spectra, following
\citet{Matt01}. The soft excess will be treated separately in Sect. \ref{softpar}
and put in the general context of the broadband model in Sect. \ref{discussion}.

\subsubsection{\label{beppobroad}The broadband BeppoSAX data}

In order to correct the residuals on the high energy part of the spectrum, we
changed our baseline model introducing a reflection component from neutral
matter \citep[\textsc{pexrav} model: ][]{mz95} and an iron line. The same model
successfully correct the residuals observed in the three BeppoSAX observations
(see Fig. \ref{saxpexrav}).

The best fit energies of the iron lines are clearly shifted to energies higher
than 6.4 keV (respectively $6.69^{+0.09}_{-0.13}$, $6.60^{+0.07}_{-0.18}$ and
$6.57^{+0.14}_{-0.07}$ keV for the three observations). Moreover, the lines are
resolved ($\simeq0.3$ keV). These results are consistent with those derived by
\citet{per02}, but, interestingly, they are marginally different from the value
of the line centroid in $ASCA$, being $6.44^{+0.06}_{-0.04}$, which in turns is
consistent with the results published by \citet{wang99}. The reason for this
discrepancy is probably due to an intrinsic complexity of the iron line of this
source combined with the different response of the two instruments at high
energies.

\citet{Matt01} found that the XMM-$Newton$ data clearly show at least two components for the iron line, one narrow at 6.4 keV and the other broad at higher energies. They suggested that the best explanation for the broad component is in
terms of a blend of ionized iron lines, Fe \textsc{xxv} at 6.7 and Fe
\textsc{xxvi} at 6.96 keV. Therefore we decided to adopt a Compton reflection
model plus three narrow gaussian lines (with fixed centroid energies) also for
the BeppoSAX data (see Table \ref{saxfits}). All the other parameters do not
change significantly from the initial model with a single, broad iron line, but
the residuals around the line are better corrected and a marginal improvement of
the $\chi^2$ is achieved. If this interpretation of the iron line complex is
correct, the $ASCA$ observation of a single neutral iron line can be tentatively explained
as due to the steeper decline of the effective area of this instrument at these energies.

\begin{table}
\caption{\label{saxfits}Compton reflection plus three gaussian lines model for
the three BeppoSAX observations (MECS+PDS). The inclination angle of the reflecting matter is fixed
to $30\degr$ (see text for details).}

\begin{tabular}{llll}
\hline 
\\
\textbf{Date}  & \textbf{01/30/1997}  & \textbf{01/14/1998} &
\textbf{02/01/2001} \\
\\
$\Gamma$& $2.08^{+0.07}_{-0.08}$ & $1.93^{+0.11}_{-0.10}$ &
$2.02^{+0.08}_{-0.07}$ \\

N$_\mathrm{H}$ ($10^{22}$ cm$^{-2}$)& $3.77^{+0.18}_{-0.19}$ &
$3.55^{+0.21}_{-0.22}$ & $3.42^{+0.24}_{-0.14}$ \\

E$_\mathrm{c}$ (keV) & $>318$ & $>169$ & $255^{+235}_{-89}$ \\

R & $1.48^{+0.44}_{-0.21}$ & $1.02^{+0.50}_{-0.38}$ & $1.51^{+0.42}_{-0.34}$\\

F$_{6.4\,\mathrm{keV}}$ ($10^{-5}$ cgs) & $7.9^{+3.3}_{-1.8}$ & $9.5^{+4.9}_{-4.9}$ &
$8.6^{+2.5}_{-5.0}$\\

EW$_{6.4\,\mathrm{keV}}$ (eV)& $83^{+35}_{-19}$ & $94^{+49}_{-48}$ & $89^{+26}_{-52}$\\

EW$_{6.7\,\mathrm{keV}}$ (eV)& $<24$ & $<120$ & $<54$\\

EW$_{6.96\,\mathrm{keV}}$ (eV)& $68^{+110}_{-42}$ & $<72$ & $50^{+70}_{-25}$\\

F$_{2-10\,\mathrm{keV}}$ (cgs) & $7.2\times10^{-11}$ & $7.1\times10^{-11}$ &
$7.8\times10^{-11}$\\

$\chi^2$/dof& 76/88 & 92/82 & 88/91 \\
\\
\hline
\end{tabular}

\end{table}

The model with Compton reflection from neutral matter and three narrow gaussian
lines is statistically very good also for the pn-PDS spectrum from the
simultaneous XMM-SAX observation (see third column in Table \ref{xionpexrav}).

\subsubsection{\label{iron}The iron line complex}

\begin{figure}
\centerline{\epsfig{figure=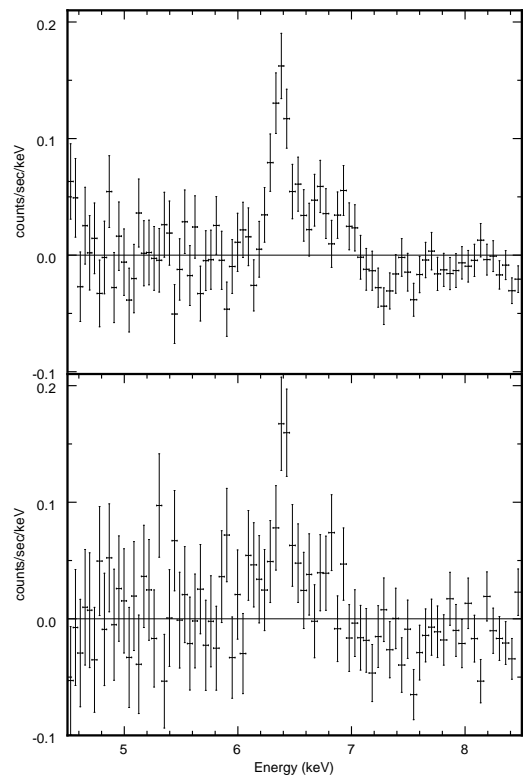,width=7cm}}
\caption{\label{iron_res}EPIC-pn residuals around the iron line: first (top) and
second (bottom) XMM observation.}
\end{figure}

As already mentioned, \citet{Matt01} found that the pn spectrum showed an iron line composed by at least two components, one narrow, at 6.4 keV, and the other broader, bluewards. This structure for the line complex is less clear in the second XMM observations, where the total flux of the source was almost doubled and the broader component is less visible (see Fig. \ref{iron_res}).

\begin{figure*}

\begin{center}

\subfigure[\label{6.4}]
{\epsfig{figure=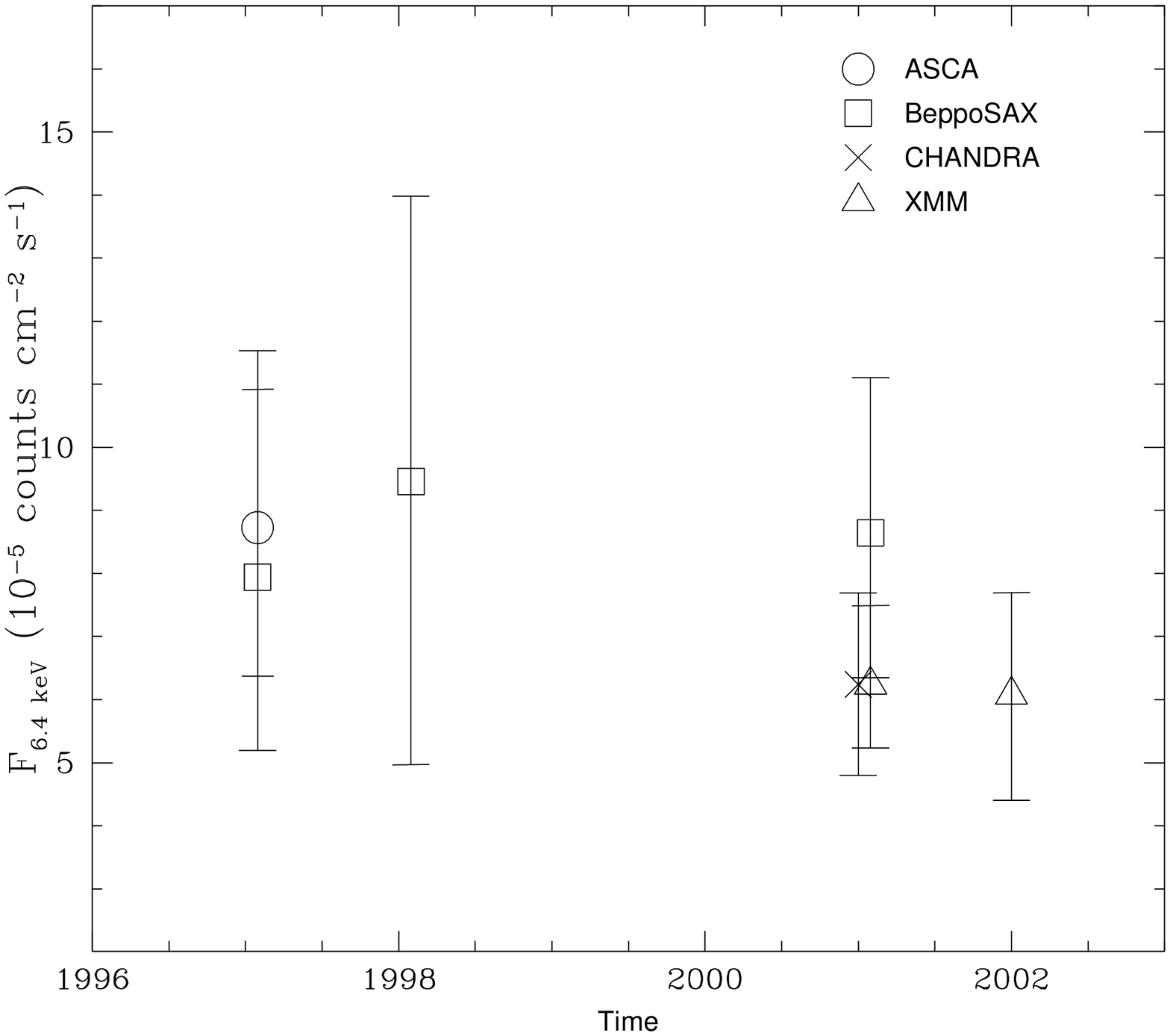,width=5.2cm}}
\subfigure[\label{6.7}]
{\epsfig{figure=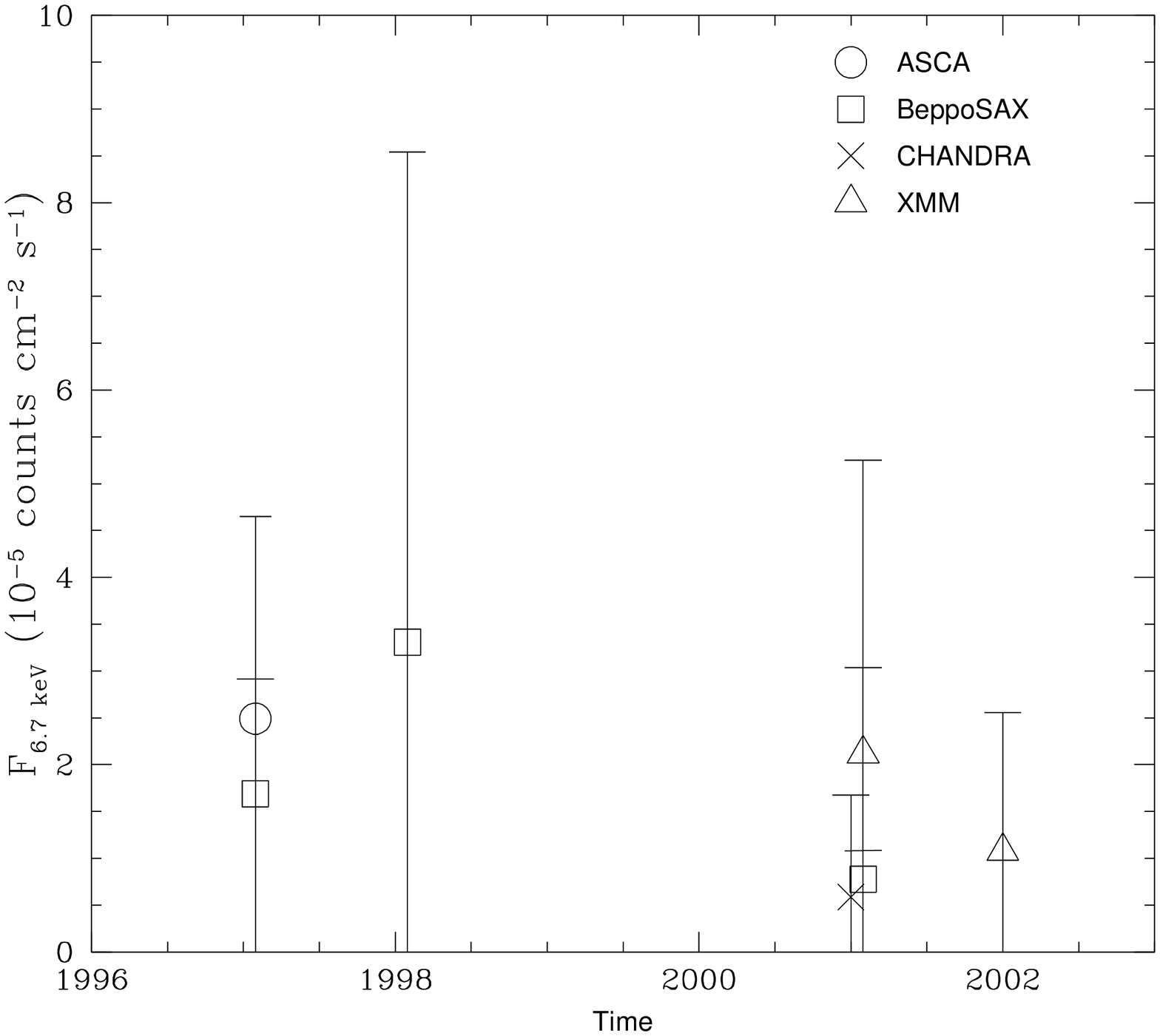,width=5.2cm}}
\subfigure[\label{6.96}]
{\epsfig{figure=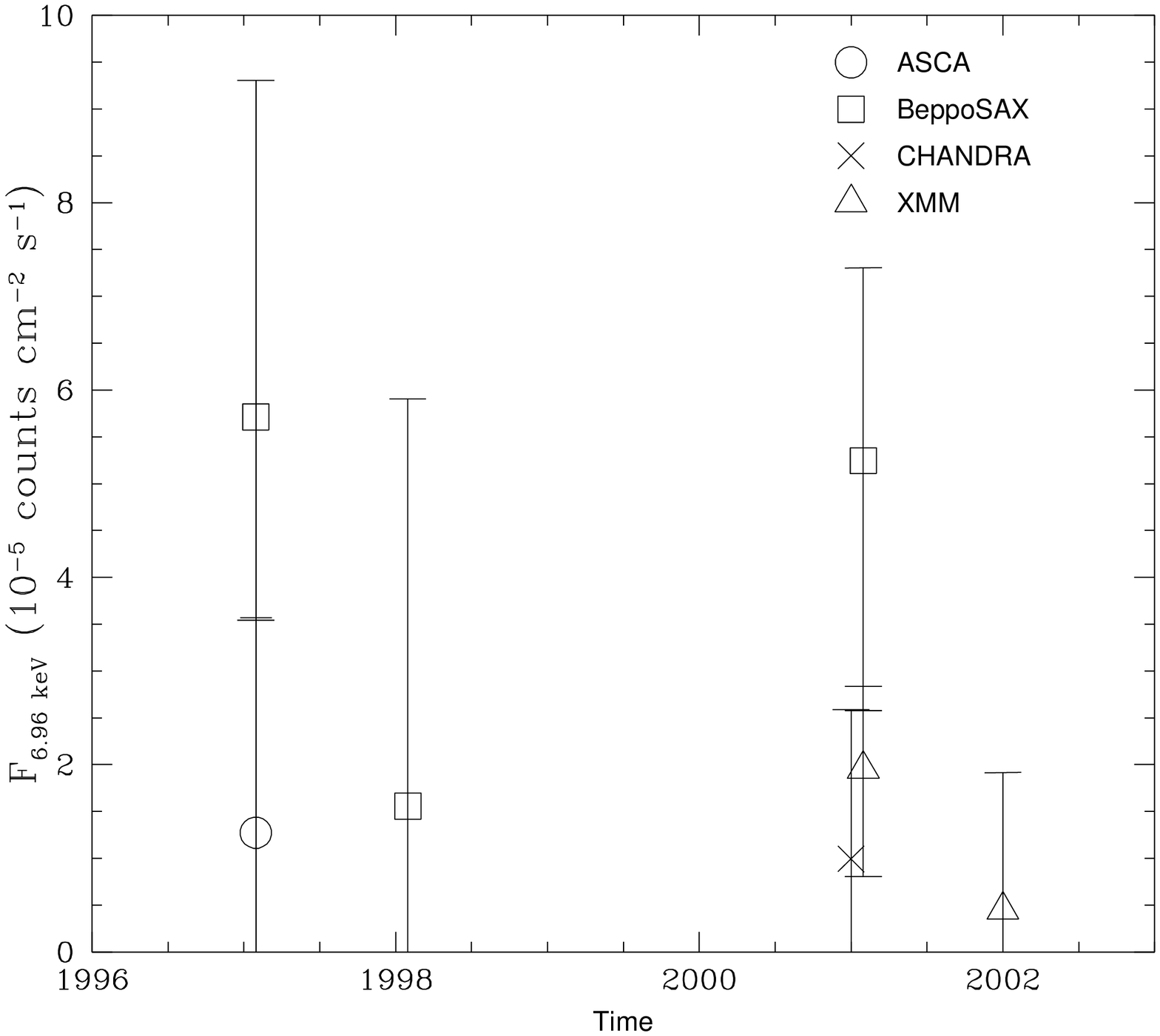,width=5.2cm}}

\end{center}

\caption{Fluxes of the iron lines in the analysed observations of NGC~5506: 6.4
(a), 6.7 (b) and 6.96 (c) keV. The centroid energies of the lines were kept
fixed in the fits (see text for details).}

\end{figure*}

The line at 6.4 keV is unresolved at the CCD resolution, in both the XMM observations. The higher spectral resolution of the gratings aboard $Chandra$ allows a tighter upper limit on the FWHM of the line, which is about 4000 km s$^{-1}$ (99\% confidence level). This value clearly excludes an origin from the innermost region of an accretion disc, suggesting instead a more distant matter, such as the BLR or the torus. This is also supported by the absence of flux
variability of the line in all the analysed observations (Fig. \ref{6.4}). Because NGC~5506 was recently discovered to be a NLS1, with BLR line widths $<$ 2000 km s$^{-1}$ \citep{nag02}, the upper limit to the line width from $Chandra$
cannot permit to exclude an origin of the line from the BLR. On the other hand, it is preferable to identify the narrow line with the one inevitably arising from the matter producing the reflection component: we shall return on this subject in Sect. \ref{broadband}. It seems then reasonable that a torus is the most likely explanation for the 6.4 keV line
and the reflection at high energies. It is worth noting that this torus does not intercept the line of sight, as the column density of the absorber is too low to explain the observed line and Compton reflection continuum \citep[see e.g.][]{mg02}. Therefore, in NGC~5506 both Compton--thin and Compton--thick circumnuclear matter are simultaneously present.

\begin{figure}

\centerline{\epsfig{figure=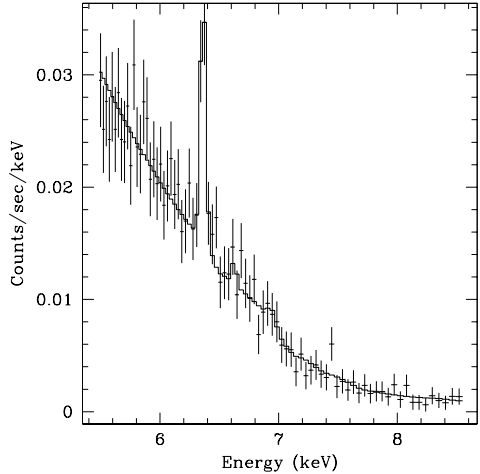,width=7cm}}
\caption{\label{iron_chandra}The $Chandra$ HEG spectrum of NGC~5506 around the
iron line. The fit (solid line) includes three gaussian lines at 6.4, 6.7 and
6.96 keV.}

\end{figure}

Following \citet{Matt01} and the model we adopted for the BeppoSAX data, we interpreted the broad component of the iron line at higher energies as the blend of two ionized iron lines at 6.7 and 6.96 keV. Those lines have equivalent widths of a few tens eV in the first XMM observation (see Table \ref{xionpexrav}) and around 10 eV in the second one, when the flux of the continuum almost doubled. Indeed, even if they are often only upper limits, the fluxes of the two ionized lines are consistent to be constant among all the analysed observations (Figs. \ref{6.7}, \ref{6.96}). The ionized iron lines are not detected in the $Chandra$ HEG spectrum (see Fig. \ref{iron_chandra}). However, if two gaussian lines are added to the fit at 6.7 and 6.96 keV, their upper limits are fully compatible with the flux measured in the XMM observations.The reason for the lack of detection in the grating spectra could be related to the physical origin of these lines and will be investigated in Sect. \ref{softpar}.

\subsubsection{\label{ionizdisc}An ionized disc?}

\begin{figure}

\centerline{\epsfig{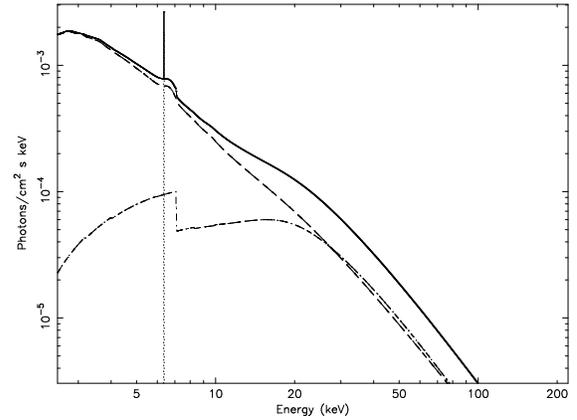}}
\caption{\label{xionmodel}A model including a cut-off power law reprocessed by an ionized disc and Compton-thick neutral matter, plus a narrow line at 6.4 keV (see text for details).}

\end{figure}

Even if the interpretation of the broad component of the iron line as a blend of two ionized lines successfully fits the data, we decided to test the alternative hypothesis that this component is actually the result of reprocessing by an ionized disc. This interpretation was tried and rejected by \cite{Matt01}, on the basis of a model including \textsc{refsch} and \textsc{diskline}. We therefore decided to adopt a more refined model, \textsc{xion} \citep{nkk_xion}, to fit the simultaneous pn/PDS spectrum. This model consists of reflection from a photoionized disc and the associated iron line, including relativistic effects. The adopted geometry is that of a lamp-post, with free parameters of the model being the height of the X-ray source above the disk, h$_\mathrm{x}$, the  accretion rate through the disc, $\mathrm{\dot{m}}$, the ratio between the luminosity of the X-ray source and that of the disc,  L$_\mathrm{x}$ / L$_\mathrm{d}$, the inner and outer disk radii (the first kept fixed to 3 R$_\mathrm{s}$ in our fits) and the spectral index.We further added a narrow line at 6.4 keV (which cannot be produced in an ionized disc) to this model. The fit is much worse than the one with reflection from neutral matter ($\Delta\chi^2=31$ with 2 less dof), mostly due to a difficulty in reproducing the neutral iron edge at 7.1 keV. Therefore, it seems likely that most (if not all) the observed Compton reflection should be produced by a Compton-thick, cold material, where the neutral iron line is also likely produced. Thus, the following step was to add a further component, a Compton reflection from neutral matter modelled as \textsc{pexrav} with R fixed to a negative value (to include only the reflected spectrum) and the spectral index linked to that of \textsc{xion} . The result is now statistically as good as the one obtained with only the Compton reflection and the three lines (see Table \ref{xionpexrav}). Fig. \ref{xionmodel} shows the components of this model: most of the reflection at high energy, the iron edge and the narrow line are produced by neutral matter, while the broad component of the line originates from the ionized disc, whose reflected continuum does not contribute too much to the high energy spectrum.

However, the unbounded values of the distance of the illuminating lamp-post from the disc and the outer radius of the disc itself in the fit (see Table \ref{xionpexrav}) are clearly indicating that there are not relativistic signatures on the spectrum. Moreover, the value of L$_\mathrm{x}$ / L$_\mathrm{d}$ is very high, requiring an extreme fine-tuning for a magnetic flares model. These results are very similar to those obtained by \citet{Matt01} with a model including \textsc{pexrav}, \textsc{refsch} and \textsc{diskline}, which is just a less sophisticated physical picture of the one we adopted: the line is not very broad and requires a large outer radius for the disc (lack of relativistic signatures) and Compton reflection from the ionized disc is negligible if compared to that from neutral matter. These considerations make the ionized disc origin of the broad component less tenable than the identification as a blend of ionized lines, despite they are statistically equivalent.

\begin{table}
\caption{\label{xionpexrav}Best fit parameters for the simultaneous pn-PDS data above 2.5 keV. The two different adopted models are described in detail in Sect. \ref{beppobroad} and \ref{ionizdisc}}

\begin{tabular}{lll}
\hline
\\
& \textbf{\textsc{xion+pexrav}}  & \textbf{\textsc{pexrav+3 gauss}}  \\
\\
N$_\mathrm{H}$ ($10^{22}$ cm$^{-2}$) & $3.96^{+0.29}_{-0.25}$ & $3.53^{+0.43}_{-0.28}$\\
$\Gamma$ & $2.05^{+0.10}_{-0.07}$ & $2.01^{+0.10}_{-0.08}$\\
E$_\mathrm{c}$ (keV) & $1000^{+1900}_{-400}$ & $220^{+490}_{-60}$ \\
R & - & $1.48^{+0.54}_{-0.36}$ \\
h$_\mathrm{x}$ (R$_\mathrm{s}$) & $>39$ & - \\
L$_\mathrm{x}$ / L$_\mathrm{d}$ & $20.0^{+1.9}_{-1.5}$ & - \\
$\mathrm{\dot{m}}$ (Eddington rate) & $<0.85$ & - \\
Outer radius (R$_\mathrm{s}$) & $>150$ & - \\
EW$_{6.4\,\mathrm{keV}}$ (eV) & $68^{+9}_{-15}$ & $88^{+20}_{-13}$\\
EW$_{6.7\,\mathrm{keV}}$ (eV) & - & $27^{+17}_{-12}$\\
EW$_{6.96\,\mathrm{keV}}$ (eV) & - & $41^{+11}_{-27}$\\
$\chi^2$/dof& 152/170 & 150/172 \\
\\

\hline

\end{tabular}

\end{table}

\subsubsection{\label{softpar}The soft excess}

\begin{figure}

\centerline{\epsfig{figure=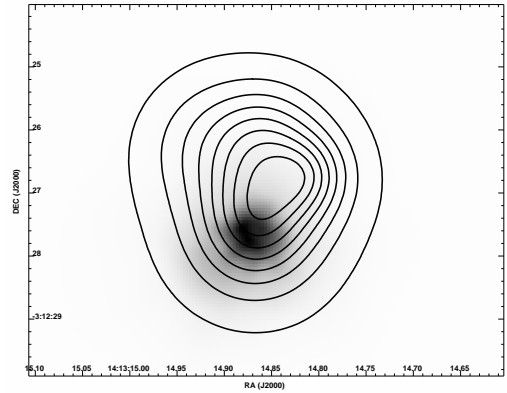,width=7cm}}

\caption{\label{soft}$Chandra$ contour levels of the emission below 1 keV
superimposed over the total emission}

\end{figure}

The $Chandra$ image of NGC~5506 clearly shows that the emission below 1 keV is extended on a radius of $\simeq350$ pc (Fig. \ref{soft}). Furthermore, this region is clearly asymmetric with respect to the nuclear emission. Interestingly the
dimension and the asymmetry of the X-ray emission seems to agree fairly well with the radio maps presented by \citet{schm01}. On the other hand, no clear evidence of a possible extension on larger scale, as claimed by \citet{colbert98} on $ROSAT$ HRI data, was found.

This extended gas is naturally associated to the soft excess best observed in the two XMM spectra and the $ASCA$ one, allowing us to exclude thermal bremsstrahlung or a blackbody as possible explanations, since it is hard to imagine such a spectral shape arising from a 350 pc scale gas. Therefore, in order to fit the residuals on the low energy part of the spectrum, we are left with two physically consistent models: reflection of the primary continuum from a photoionized gas or thermal emission from diffuse gas (\textsc{mekal}), which was found to give good fits in a sample of Compton--thin Seyfert 2s observed by BeppoSAX \citep{risa02}. The first one is parametrized (when self absorption effects are neglected) by a power law with the spectral index linked to that of the primary \textsc{pexrav} continuum responsible for the photoionization of the gas itself. We further allowed for absorption by an intervening column density of neutral gas. The best fits for the two models are presented in Table \ref{softfit}: even if marginally better from a statistical point of view, the \textsc{mekal} fits have metal abundances unphysically low. This means that the resulting spectrum lacks emission lines and is virtually indistinguishable from the already rejected bremsstrahlung interpretation. Reflection from photoionized gas remains then the only tenable model. The observed intervening absorption is $\simeq2\times 10^{21}$ cm$^{-2}$, in excess of the Galactic one, and is possibly due to the host galaxy, which is seen almost edge--on.

\begin{table}
\caption{\label{softfit}Best fit parameters for the models adopted to explain the soft excess in the two XMM observations (pn+MOS1+MOS2) and the $ASCA$ one. The fluxes are relative to the power law model.}

\begin{center}
\begin{tabular}{llll}
&\textbf{ASCA97} & \textbf{XMM01} & \textbf{XMM02}\\
\hline
\textbf{\textsc{powerlaw}}&&&\\
$\chi^2$/dof & $702/702$ & $674/611$ & $727/611$\\
N$_\mathrm{Hs}$ ($10^{21}$ cm$^{-2}$)& $1.8^{+0.3}_{-0.7}$ &
$1.5^{+0.2}_{-0.1}$ & $1.7^{+0.2}_{-0.2}$\\
&&&\\
\hline
\textbf{\textsc{mekal}} &&&\\
$\chi^2$/dof & $689/700$ & $665/609$ & $717/609$\\
N$_\mathrm{Hs}$ ($10^{21}$ cm$^{-2}$)& $2.8^{+2.3}_{-0.8}$ &
$1.9^{+0.3}_{-0.3}$ &$2.1^{+0.5}_{-0.5}$\\
kT (keV)& $0.73^{+0.14}_{-0.17}$ & $0.80^{+0.20}_{-0.13}$ &
$0.68^{+0.15}_{-0.12}$\\
A$_Z$& $0.03^{+0.07}_{-0.02}$ &$<0.02$ & $<0.02$ \\
&&&\\
\hline
\textbf{Flux 0.5-1 keV}&&&\\
($10^{-13}$ cgs)& $1.80\pm0.01$ & $2.14\pm0.01$ & $2.22\pm0.01$\\
\end{tabular}
\end{center}

\end{table}

As expected from an origin from diffuse gas, the flux of this component is remarkably constant between the two XMM
observations (see Table \ref{softfit}), while the flux of the primary component almost doubles (see Sect. \ref{data}). Therefore, the ratio of the soft excess to the nuclear flux varies from 2 to 1 $\%$ in the two observations. If the
interpretation as reflection from photoionized matter is correct, then the column density of this gas should be a few times $10^{22}$ cm$^{-2}$.  \citet{bm02} have calculate the equivalent widths of the ionized iron lines produced in gas photoionized by a power law continuum. They found that equivalent widths (with respect to the total continuum, primary plus reflected) of the order of tens of eV can indeed be produced by a photoionized gas with a column density around $10^{22}$ cm$^{-2}$ and a modest iron overabundance.  

In order to understand if our model of reflection from photoionized matter is still tenable when the gas is distributed on the observed scale, we performed a test with \textsc{cloudy}. The density of this gas was assumed to have a $r^{-2}$ behaviour, normalized in order to produce the inferred column density of $\simeq10^{22}$ cm$^{-2}$ at a radius of $\simeq350$ pc. The incident continuum was taken with the 2-10 keV observed mean luminosity and spectral shape. We have found a solution with an ionization structure fully compatible with the production of ionized iron lines \citep[see][ for details]{bm02}, which does not require a temperature higher than $10^5$ K. Such temperatures make thermal emission from the gas completely
negligible, so the model is self--consistent. Moreover, if the gas where they are emitted is spatially extended, the dispersion by the gratings would become rather complicated making the ionized lines difficult to detect: this would be a reason for the lack of their detection in the $Chandra$ HEG spectrum (see Sect. \ref{iron}). Unfortunately, the statistic is too poor to directly search for extended line emission in the $Chandra$ images.

\section{\label{discussion}Discussion}

In this section we summarize the results on all the components of the X-ray spectrum of NGC~5506 and put them in the context of a general scenario.

\subsection{\label{broadband}The broadband model}

The central engine is a Seyfert 1 nucleus, whose spectrum is as usual
represented by a power law with an exponential cutoff. The nuclear emission is
reflected from a Compton-thick neutral material, producing the characteristic
`bump' at high energies \citep[e.g. ][]{mpp91}. The narrow 6.4 keV iron line and
the 7.1 keV edge from neutral iron are likely to originate from the same matter. The value of the equivalent width of the line is fully compatible with the amount of the Compton reflection. However, a possible iron underabundance is suggested by the data (see table \ref{broadfit}): an EW of about 135 eV would be expected if R=0.75, $\theta=30\degr$ and $\Gamma=1.9$ \citep{gf91,mpp91}. In particular, the line, being narrow and keeping a constant flux despite large
variation in the total flux of the source, constrains this material to be fairly distant from the black hole. This makes a parsec scale torus the best identification for this gas. It is interesting to note that the only evidence for the
presence of a Compton-thick torus in NGC~5506 is indirect, being the signatures
of reflection of the nuclear radiation on its inner walls.

The direct nuclear emission and the reflection components are obscured by
another neutral material, with a column density of N$_\mathrm{Hc}\simeq3\times10^{22}$
cm$^{-2}$. The presence of Compton-thick and Compton-thin matter in the
circumnuclear regions of the same source seems to be a widespread phenomenon in
the environment of Seyfert galaxy \citep[see e.g.][ and references
therein]{mg02}. In this respect, NGC~5506 is yet another example.

Extended on a region of 350 pc radii, a gas, photoionized by the nuclear radiation, surrounds the nucleus.
The primary continuum is simply reflected, producing a component which
is again a power law with the same photon index as the incident spectrum. The
total column density of this gas is $\simeq10^{22}$ cm$^{-2}$, calculated
from the ratio between the normalizations of the reflected and the nuclear
continuum. The ionization structure of this gas is such that ionized iron lines
at 6.7 and 6.96 keV are produced, with equivalent widths against the total
continuum of tens of eV. The radiation reflected from this gas is further
obscured by a neutral gas along the line of sight with column density of
N$_\mathrm{Hs}\simeq2\times10^{21}$ cm$^{-2}$, in excess to that of our Galaxy. This gas
could naturally be associated with the host galaxy, which is seen almost
edge-on.

The best fit parameters for this model are shown in Table \ref{broadfit}, when
used on the pn-MOS-PDS and the SIS-PDS spectra. The results for the two sets of
simultaneous observations are consistent with each other, with the only
exception of the photon index and the column density N$_\mathrm{Hc}$. The
marginal inconsistency of these values could be partly due to the fact that the
MOS spectra were fitted down to 0.3 keV, while the SIS ones to 0.5 keV,
introducing some discrepancies in the determination of the spectral index and
the cold absorption.

\begin{table}
\caption{\label{broadfit}Best fit parameters for the broadband model adopted
for NGC~5506 on the pn-MOS-PDS and the SIS-PDS spectra (see text for details).}

\begin{center}
\begin{tabular}{lcc}
\hline
&\\
&\textbf{pn-MOS-PDS (2001)} & \textbf{SIS-PDS (1997)}\\
&\\
$\Gamma$ & $1.81^{+0.03}_{-0.06}$ & $1.94^{+0.10}_{-0.08}$\\
E$_\mathrm{c}$ (keV) & $130^{+35}_{-40}$ & $150^{+140}_{-50}$\\
R & $0.75^{+0.15}_{-0.11}$ & $1.4^{+0.8}_{-0.5}$\\
N$_\mathrm{Hc}$ ($10^{22}$ cm$^{-2}$) & $2.89^{+0.02}_{-0.04}$ & $3.2^{+0.1}_{-0.1}$\\
N$_\mathrm{Hs}$ ($10^{21}$ cm$^{-2}$) & $1.52^{+0.13}_{-0.19}$ & $1.9^{+0.8}_{-0.8}$\\
EW$_{6.4\,\mathrm{keV}}$ (eV) & $86^{+22}_{-10}$ & $88^{+24}_{-32}$\\
EW$_{6.7\,\mathrm{keV}}$ (eV) & $26^{+9}_{-9}$ & $30^{+19}_{-24}$\\
EW$_{6.96\,\mathrm{keV}}$ (eV) & $28^{+10}_{-11}$ & $<24$\\
F$_{2-10\,\mathrm{keV}}$ (cgs) & $5.8\times10^{-11}$ & $6.9\times10^{-11}$\\
$\chi^2$/dof & $709/627$ & $441/438$\\
&\\
\hline
\end{tabular}
\end{center}

\end{table}

\begin{figure}

\centerline{\epsfig{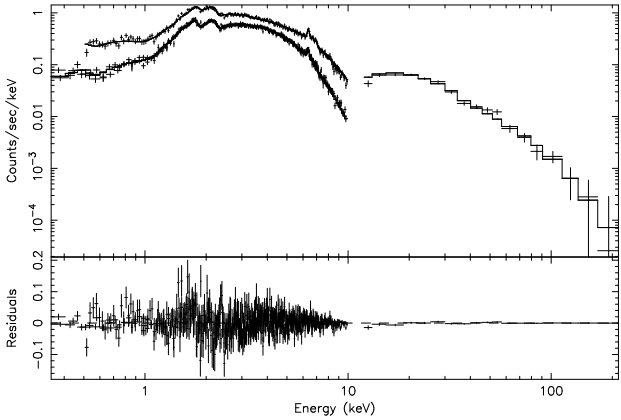}}
\caption{\label{pnmos12}Best fit model for the pn-MOS-PDS data (see text and Table \ref{broadfit}for details).}

\end{figure}

\subsection{Where is the accretion disc?}

It is worth noting that our model does not contain any component directly related to the accretion disc. Indeed, the results from the fits with the model \textsc{xion} (see Sect. \ref{ionizdisc}) require quite extreme conditions and basically confirm that there is no signature of reprocessing from the accretion disc in our data. There are at least two possible explanations to this problem.

It is possible that the accretion disc is completely ionized. In this case, the reflection component would be almost indistinguishable from the primary continuum, being a simple power law. Moreover, neither iron lines nor edges
would be produced because all the elements would be completely stripped off, leaving no signatures for the presence of the disc, at least in the energy band the cold absorption present in NGC~5506 let us investigate (i.e. above a few
keV). Very ionized accretion discs are generally produced by high accretion rates \citep[e.g.][ and references therein]{ball01}. It is possible to estimate the accretion rate for NGC~5506 from the mass of the central BH, as inferred from \citet{hay98} from X-ray variability. Even if their value is highly model dependent, a mass of $2\times10^6$ M$_{\sun}$ would produce an Eddington luminosity of $3\times10^{44}$ erg s$^{-1}$. Starting from the mean unabsorbed 2-10 keV X-ray luminosity we observed ($\approx6\times10^{42}$ erg s$^{-1}$), we can derive the bolometric luminosity of the source by multiplying by a factor 10-30 \citep{elvis94}. Therefore, the Eddington ratio for NGC~5506 is $\eta\approx0.2-0.6$, which in turns means a
high accretion rate. Low mass BHs and high values of the Eddington ratio are generally believed to be typical characteristics of NLS1s, as physical explanations for the spectral peculiarities of these sources \citep[see e.g.][]{bbf96,laor97,boroson02}. In this respect, ionized discs could be ubiquitous in NLS1s as a result of the required high accretion rate and recently
there have been claims of their presence in these sources \citep[see][ and references therein]{ball01}. 

A second solution would be that the accretion disc is seen almost edge-on, making the reprocessed components hardly detectable because of their low intensity \citep[see e.g.][]{mpp91}. This solution, even if not implausible, is not favoured by the fact that the torus clearly does not intercept the line of sight. Therefore, a geometry would be required with the disc and the
torus which are not co-aligned as would be, at least naively, expected. Interestingly, such a scenario is required in another NLS1, RE~J1034+396, whose inferred values of BH mass and Eddington ratio are of the same order as NGC~5506: this source is best modelled by an almost edge-on disc, while no torus again intercepts the line of sight \citep{puch01}.

\section{Conclusions}

The analysis of all the data available from XMM-$Newton$, $Chandra$, BeppoSAX and $ASCA$ led us to picture a self-consistent scenario for the X-ray emission of the NLS1 NGC~5506:
\begin{itemize}
\item Both Compton--thin and Compton--thick circumnuclear matter are simultaneously present: the former intercepts the line of sight and absorbs the source flux at soft energies; the latter is indirectly observed through the signatures left in the spectrum, a Compton reflection component, a neutral iron edge and a narrow K$\alpha$ line
\item The iron line consists of a narrow neutral component, arising from the torus, and a blend of ionized iron lines at 6.7 and 6.96 keV
\item The nucleus is surrounded by an extended ($\approx350$ pc) soft emission, likely associated with reflection of
the nuclear emission by a photoionized gas, where both the soft excess and the ionized iron lines are produced
\item There is no spectral signature for the presence of the accretion disc: possible solutions could be a fully ionized or almost edge-on disc
\end{itemize}  

\acknowledgement

SB, IB, GM and GCP acknowledge ASI and MIUR (under grant \textsc{cofin}-00-02-36) for financial support. We would like to thank Sergei Nayakshin for his helpful comments on the \textsc{xion} code, and the anonymous referee.

\bibliographystyle{aa}
\bibliography{sbs}

\end{document}